\documentclass[journal ]{new-aiaa}
\usepackage[utf8]{inputenc}
\usepackage{textcomp}

\usepackage{graphicx}
\usepackage{amsmath}
\usepackage[version=4]{mhchem}
\usepackage{siunitx}
\usepackage{longtable,tabularx}
\usepackage{multirow}
\usepackage{svg}

\title{High specific impulse electrospray propulsion with small capillary emitters}

\author{Manel Caballero-P\'erez \footnote{PhD student, Mechanical and Aerospace Engineering, 4200 Engineering Gateway}, Marc Galobardes-Esteban \footnote{PhD candidate, Mechanical and Aerospace Engineering, 4200 Engineering Gateway}, and Manuel Gamero-Casta\~no \footnote{Professor, Mechanical and Aerospace Engineering, 4200 Engineering Gateway, AIAA Senior Member}}
\affil{University of California, Irvine, Irvine, CA, 92617}

\begin{document}

\maketitle

\begin{abstract}

This study demonstrates the feasibility of using smaller capillary emitters to achieve higher specific impulse ($I_\text{sp}$) in electrospray propulsion. Four ionic liquids were characterized using capillary emitters with tip diameters from 15 to 50\,\textmu m. Smaller diameter capillaries produced smaller and more stable Taylor cones. This stabilization enabled steady cone-jet operation at significantly lower flow rates compared to larger emitters. This was unexpected because when the jet diameter is much smaller than far-field geometric features, the minimum flow rate is thought to be solely determined by the physical properties of the propellant. Using the smaller emitters and acceleration voltages of 10\,kV, specific impulses up to 3000\,s could be achieved with efficiencies above 50\%, approximately doubling the $I_\text{sp}$ observed with larger emitters. For one of the liquids and the smallest emitters, the beam consisted solely of ions at the lowest flow rates, similarly to studies using externally wetted and porous emitters. Another important finding was that at sufficiently low flow rates, a significant fraction of the propellant fed to the emitter is not accelerated by the electrostatic field. These propellant losses make the time-of-flight technique unreliable for determining the $I_\text{sp}$.   

\end{abstract}

\section{Introduction}
\lettrine{E}{lectrospray} propulsion relies on the electrostatic acceleration of ions and charged droplets emitted from an electrospray source. An electrospray emitter produces sub-micronewton level thrust while operating at relatively high specific impulse ($I_\text{sp}$) and propulsive efficiency \cite{gamero2001electrospray}. To generate adequate thrust for satellites, arrays comprising hundreds or thousands of individual emitters are constructed using microfabrication techniques \cite{lenguito2013development,grustan2017microfabricated},  conventional machining \cite{natisin2020fabrication,dworski2024investigating}, or other methods \cite{kim2025high}. The concept of electrospray propulsion was introduced in the 1960s, when conductive organic liquids and liquid metals were tested \cite{krohn1961liquid,krohn1963glycerol}. These tests were not very successful, as the organic liquids produced droplets with very high mass-to-charge ratios leading to insufficient $I_\text{sp}$, while liquid metals emitted large numbers of ions along with droplets. However, the need for efficient micropropulsion reignited interest in this technology in the 1990s. Since then, the term electrospray propulsion (ESP) has been used for systems employing organic propellants, while field emission electric propulsion (FEEP) refers to those using liquid metals \cite{martinezsanchez1999,marcuccio1998experimental}. 

FEEP thrusters have notable flight heritage \cite{krejci2024informing}, but they produce relatively low thrust due to relatively low propulsive efficiencies (\SIrange{10}{15}{\percent} \cite{krejci2019full}) and high operating $I_\text{sp}$ (\SIrange{1500}{5000}{\second})\footnotemark[1]. These characteristics are disadvantageous for missions requiring higher thrust-to-power ratios, such as those in a high drag environment or involving time-sensitive maneuvers like orbit raising. Additionally, FEEP thrusters require heating of the propellant, which can consume 10\% or more of the thruster's total power\footnotemark[1]. In contrast, most ESP thrusters do not require heating and offer the flexibility to operate efficiently in either ion-dominated or droplet-dominated regimes. Ionic liquids are commonly used as propellants in ESP because their low vapor pressures make them suitable for operation in space. Alternative propellants, such as electrolytes of salts dissolved in low-vapor-pressure solvents, have been considered but tend to suffer from excessive mass loss due to evaporation \cite{garoz2013charge}. Some protic ionic liquids, such as methylammonium formate, also experience significant evaporation losses owing to their higher vapor pressures \cite{garoz2006taylor,garoz2007taylor} compared to typical aprotic ionic liquids used in electrospray thrusters, like 1-ethyl-3-methylimidazolium bis(trifluoromethylsulfonyl)imide (EMI-Im) or 1-ethyl-3-methylimidazolium tetrafluoroborate (EMI-BF$_4$). Droplet-dominated regimes are associated with higher thrust levels and relatively lower $I_\text{sp}$, while ion-dominated regimes often achieve $I_\text{sp}$ values around \SI{1500}{\second} at acceleration voltages of \SIrange{1}{2}{\kilo\volt}, albeit with lower thrust \cite{villegas2024emission,huang2023performance}. This versatility makes ionic liquids particularly attractive for a wide range of mission profiles. This study focuses exclusively on electrosprays of ionic liquids.  \footnotetext[1]{ENPULSION NANO-R\textsuperscript{3} Product Overview, URL: https://www.enpulsion.com/wp-content/uploads/ENP2019-086.F-ENPULSION-NANO-R\%C2\%B3-Product-Overview.pdf} 
    
Electrospray devices can be classified based on emitter geometry into three types: internally fed (capillary emitter) electrosprays, externally wetted emitter electrosprays, and porous emitter sprays. In capillary emitter electrosprays, the liquid is actively fed toward the emitter tip by controlling the pressure at the propellant reservoir. A high voltage is applied between the liquid and a grounded extractor electrode located in front of the emitter. The intense electric field deforms the liquid meniscus at the capillary tip into a cone, commonly known as a Taylor cone, from whose apex a thin jet is ejected. The base diameter of the Taylor cone is determined by the external diameter at the capillary tip (typically in the range of \SIrange{30}{100}{\micro\meter} \cite{cisquella2022scalable,lenguito2013pressure}) for usual liquid-emitter contact angles, and has a half-angle of approximately \ang{49} \cite{fernandez2007fluid}. The jet eventually breaks up due to the Rayleigh capillary instability, forming a beam of charged droplets and molecular ions. Usually, droplet-dominated regimes are associated with capillary emitters. Externally wetted and porous emitter electrosprays, on the other hand, rely on capillary action and electrostatic suction to supply liquid to the emitter tip. The extracted flow rate can be controlled by adjusting the voltage difference between the emitter and the extractor. Unlike capillary emitter electrosprays, they typically operate in ion-dominated regimes, generally resulting in higher $I_\text{sp}$. The menisci formed in these geometries are much smaller than those in capillary emitters, with radii of curvature at the emitter tips on the order of \SIrange{5}{15}{\micro\meter} \cite{velasquez2006planar,huang2022emission}. Sometimes, wedge-shaped emitters are used instead of needle-like ones, allowing multiple emission sites to form along their edges \cite{courtney2019high,wright2021multiplexed}. A capillary emitter ESP thruster was employed in the ST7-DRS mission, achieving over 3000 hours of successful operation \cite{ziemer2017colloid, ziemer2019lisa}. In comparison, ESP devices using externally wetted or porous emitters generally exhibit shorter periods of stability \cite{castro2009effect, krejci2017emission} and may require frequent polarity switching to mitigate the effects of electrochemical reactions \cite{lozano2004ionic}. Capillary emitters offer direct control over the flow rate and protect the liquid from direct exposure to space radiation, which could compromise the ionic liquid \cite{ziemer2003colloid}. However, externally wetted and porous emitters have advantages such as a higher emitter density \cite{inoue2019fabrication}, and the potential elimination of components like valves and propellant pressurization systems, both of which can introduce added complexity to the thruster system. Additionally, capillary emitters require methods to increase hydraulic impedance of the fluid supply such as microfabricated channels \cite{grustan2017microfabricated}, whereas porous and externally wetted emitters inherently possess high hydraulic impedance.

This study investigates electrosprays of ionic liquids operating in the cone-jet mode using smaller diameter capillary emitters than those employed in previous works \cite{cisquella2022scalable, grustan2017microfabricated}. Our goal is to increase $I_\text{sp}$ while retaining the advantages of capillary emitters over externally wetted and porous emitters. The hypothesis is that smaller Taylor cones have a lower minimum flow rate at which they can be operated stably, thereby increasing the maximum $I_\text{sp}$. Capillaries with tip diameters of 15, 20, 30, 40, and \SI{50}{\micro\meter} are used, with the smallest diameters expected to generate Taylor cones comparable in size to those typical of porous or externally wetted emitters. We characterize four different ionic liquids, namely EMI-Im, 1-ethyl-3-methylimidazolium trifluoroacetate (EMI-TFA), ethylammonium nitrate (EAN), and 1-butyl-3-methylimidazolium tricyanomethanide (BMI-TCM), with each emitter size and across a range of flow rates. Table \ref{tab:properties} lists their relevant physical properties including the electrical conductivity $K$, the surface tension $\gamma$, the viscosity $\mu$, the density $\rho$, the dielectric constant $\varepsilon$ and the molecular mass of the cation ($M^+$) and anion ($M^-$). By exploring different capillary sizes, we observe significant and unexpected variations in the minimum flow rate at which electrosprays remain stable, and the continuous transition between the droplet-dominated and the ion-dominated regimes. We discuss potential reasons behind these observations and provide context based on current understanding of the physics of cone-jets. 

\begin{table}[hbt!]
\caption{\label{tab:properties} Ionic liquids investigated, physical properties at 298 K.}
\centering
\begin{tabular}{lccccccc}
    \hline
    \hline
    {} & $K$  & $\gamma$  & $\mu$  & $\rho$  & $\varepsilon$ & $M^+$  & $M^-$ \\
    Short name &  (S/m) &  (mN/m) & (mPa$\cdot$s) & (g/cm$^3$) &  &  (Da) &  (Da)\\
    \hline
    EMI-Im   & 0.92 & 36.0 & 39.6 & 1.52 & 13.8 & 111.2 & 280.1 \\
    EMI-TFA  & 0.96 & 49.0$^{\dag}$ & 33.0 & 1.29 & 14$^{\ddagger}$ & 111.2 & 113.0 \\
    EAN      & 2.27 & 48.1 & 39.8 & 1.21 & 29   & 46.1  & 62.0  \\
    BMI-TCM  & 1.03 & 49.6 & 25.7 & 1.05 & 14$^{\ddagger}$ & 139.2 & 90.1  \\
    \hline
    \hline
\end{tabular}
\vspace{2mm}

{\footnotesize
All values from \cite{dong2006ionic} unless otherwise specified; $^\dag$ from \cite{zaitsau2018imidazolium}; $^\ddagger$ unavailable in the literature, estimated from dielectric constants (12-17) of ionic liquids with imidazolium-based cations~\cite{rybinska2018prediction}.  
}
\end{table}

\subsection{Review of cone-jet electrospray physics}

 Four distinct regions can be identified within a cone-jet: the upstream meniscus or Taylor cone, the transition region bridging the cone and the jet, the jet, and the jet breakup. An additional area, termed the acceleration region, extends from the breakup point until the species exit through the extractor. Ions may be emitted from the transition region, the jet, and/or droplets \cite{gamero2000direct, gamero2021electrosprays}. In the cone region, charge transport occurs via ohmic conduction with negligible voltage drop, whereas surface convection of charge dominates in the jet \cite{gamero2019numerical}. In the transition region, the transport mechanism evolves from conduction to convection, and both the normal and tangential components of the electric field at the surface reach local maxima; additional local maxima occur in the breakup region, both on the surface of droplets and at the pinching of the jet. These local maxima of the electric field determine the regions from which ions are emitted \cite{gamero2000direct}. The ejected current is determined in the transition region, and processes taking place in this region likely control the minimum flow rate \cite{ponce2018steady, gamero2019minimum}. When ion emission either occurs in the jet breakup region, from the droplets during flight, or does not occur at all, the emitted current $I$ follows the trend \cite{ganan2018review}:

\begin{equation}
    I \cong 2.6 \left( \frac{\gamma K \dot{m} }{\rho}\right)^{1/2}
    \label{eq:curr_scaling}
\end{equation}
where $\dot{m}$ is the mass flow rate. Note that the average charge-to-mass ratio of the droplets is then given by
\begin{equation}
    \left\langle \frac{q}{m} \right\rangle  \cong  \frac{I}{\dot{m}} =   2.6 \left( \frac{\gamma K  }{\rho \dot{m}}\right)^{1/2}
    \label{eq:<q/m>}
\end{equation}
which illustrates the need to reduce the mass flow rate to increase the specific impulse, $I_{sp} \cong c/g_0 \propto \dot{m}^{-1/4} $ where $c$ is the exhaust velocity and $g_0 = \SI{9.81}{\meter/\second\squared}$. The diameters of the jet and droplets scale with $r_G$ \cite{ganan1997cone}:

\begin{equation}
    r_G =\left( \frac{\rho \varepsilon_0 Q^3}{\gamma K} \right)^{1/6} 
    \label{eq:r_G}
\end{equation}
where $\varepsilon_0$ is the vacuum permittivity and $Q$ is the volumetric flow rate. The state of a cone-jet can be parametrized with three dimensionless number, typically the dielectric constant, the dimensionless flow rate $\Pi$ and the electrohydrodynamic Reynolds number $Re_K$: 

\begin{equation}
   \Pi = \frac{\rho K Q }{\gamma \varepsilon_0}, \qquad  Re_K = \left(\frac{\gamma^2 \rho \varepsilon_0}{\mu^3 K}\right)^{1/3}
    \label{eq:Pi}
\end{equation}
The length of the transition region $H$ can be defined in several ways. One approach considers where the surface current at the cone-jet changes from 5\% to 95\% of its final value. Using this definition, a numerical study found the length to be \cite{gamero2019numerical}:

\begin{equation}
    H \approx \beta \, \Pi^\alpha \, r_G
    \label{eq:H_G}
\end{equation}
where $\alpha$ and $\beta$ are fitting parameters that primarily depend on the dielectric constant. For $\varepsilon = 8.91$, a value close to that of the ionic liquids tested in this study,  $\alpha \approx 0.17$ and $\beta \approx 16.3$ \cite{gamero2019numerical}. 

The cone-jet operates stably within a specific voltage and flow rate parameter space \cite{cloupeau1989electrostatic}. Below a certain flow rate threshold, instabilities lead to intermittent emission or dripping; significantly below this threshold, cone-jets do not form. The causes of the minimum flow rate have been rationalized in different ways. First, the minimum flow rate has been explained as an instability in the transition region caused by the balance of different stresses with the always dominant electric stress \cite{ganan2013minimum, Higuera_2017}. For example, \cite{ganan2013minimum} proposes that for liquids with $\varepsilon Re_K \ll 1$ viscous stresses balance the pressure gradient, leading to

    \begin{equation}
        \Pi_{\text{min}} \sim Re_K^{-1}
        \label{eq:minflow_visc}
    \end{equation}
while polarization forces balance the pressure gradient when $\varepsilon Re_K \gg 1$, leading to 
    \begin{equation}
        \Pi_{\text{min}} \sim \varepsilon
        \label{eq:minflow_pol}
    \end{equation}
Criteria \eqref{eq:minflow_visc} and \eqref{eq:minflow_pol} are referred to as the viscous and polarization minimum flow rate limits, respectively. 
On the other hand, reference \cite{gamero2019minimum} finds that Eq. \eqref{eq:minflow_visc} also reproduces the minimum flow rate of liquids with $\varepsilon Re_K \gg 1$, and attributes the onset of the minimum flow rate to excessive viscous dissipation at the base of the jet.

Another instability mechanism that may lead to the minimum flow rate originates from the jet's absolute instability \cite{lopez2010absolute}. Below this threshold, perturbations that would normally be convected downstream propagate upstream, destabilizing the entire flow structure. The convective and absolute instability regions are defined by the dimensionless numbers $Re_j = (2/\pi) \Pi^{1/2} Re_K$ and $Ca_j = (4/\pi) Re_K^{-1}$. The high conductivity of ionic liquid propellants leads to $Ca_j \gg 1$. In this range, the boundary between convective and absolute instability regions can be approximated by $\ln Re_j \approx -\ln Ca_j - 2$ \cite{lopez2010absolute}. When expressed in terms of the dimensionless flow rate, the boundary yields a value much smaller than one, $\Pi_\text{min}\ll 1$.

\begin{table}[hbt!]
\caption{\label{tab:tempincrease} Parameters for estimating the temperature increase along the cone-jet, Eq. \eqref{eq:tempincrease}:  
(a) $\Delta T$ at current crossover;  
(b) $\Delta T$ at \(500 \times r_G\) from the cone vertex \cite{magnani2025}.}
\centering
\begin{tabular}{llcccc}
    \hline
    \hline
    {} & {} & EMI Im & EMI TFA & BMI TCM & EAN \\ 
    \hline
    \multirow{3}{*}{(a)} 
      &$b_1$ (K$\cdot$s/kg)& 0.526 & 0.308 & 0.213 & 0.479 \\
      & $b_2$ & -0.400 & -0.420 & -0.259 & -0.243 \\
      &$b_3$ (K)& -7.63 & -8.14 & -28.2 & -39.2 \\ 
    \hline
    \multirow{3}{*}{(b)} 
      &$b_1$ (K$\cdot$s/kg)& 0.0104 & 0.00599 & 0.130 & 1.27 \\
      & $b_2$ & -0.390 & -0.414 & -0.292 & -0.223 \\
      &$b_3$ (K)& -3.61 & -5.43 & -10.2 & -52.9 \\ 
    \hline
    \hline
\end{tabular}
\end{table}

Full charge separation is another potential mechanism determining the minimum flow rate. Originally proposed by \cite{fernandez2007fluid} for dilute electrolytes, we adapt it here for ionic liquids. The liquid flows toward the meniscus, carrying cations and anions at rates $\dot{n}^+ = \dot{n}^- = \dot{n}$. Not all these ions are free; some are bound as neutral ion pairs or larger neutral clusters. The extent of ion dissociation, known as the ionicity $\nu$, is approximately 0.7 for most ionic liquids \cite{macfarlane2017fundamentals}. The emitted current is $I = \dot{n}_f^+ e$, where $\dot{n}_f^+ = \dot{n}_f^- = \dot{n}_f$ is the rate of free cations and anions, and $e$ is the elementary charge. The ratio of charge-carrying ions to total ions must be below the ionicity, $\dot{n}_f/\dot{n} \leq \nu$. For a given current, the minimum flow rate associated with full charge separation is

\begin{equation}
    Q_{\text{min}} = \frac{\dot{n}^+ M^+ + \dot{n}^- M^-}{\rho} = \frac{\dot{n}_f ( M^+ + M^-)}{\rho \nu} = I \,\frac{  M^+ + M^-}{ \rho \nu e}
    \label{eq:minflow_iontransport}
\end{equation}

Combining equations  \eqref{eq:minflow_iontransport} and \eqref{eq:curr_scaling} leads to
\begin{equation}
    \Pi_{\text{min}} = \left(\frac{  M^+ + M^-}{ \nu e}\right)^2 \frac{2.6^2  K^2}{\varepsilon_0 \rho}
    \label{eq:minflow_iontransport-2}
\end{equation}

The jet diameter is much smaller than the emitter diameter $D_e$ in most situations. In this case, it is generally accepted that geometric features play a minor role in electrosprays \cite{fernandez2007fluid, ganan2018review}, negligibly affecting the current and the minimum flow rate. This understanding is based on the premise that the electric fields governing the flow are determined solely by the local charge distribution and are unaffected by far-field features. Experimental studies have corroborated this, finding no changes in the minimum flow rate when varying the emitter diameter as long as $D_e/r_G > 10^2$. However, when these ratios are smaller than approximately $80$, smaller emitter diameters have been observed to stabilize electrosprays and lower the minimum flow rate \cite{ponce2018steady}. In mass spectrometry applications, capillaries with diameters around 100\,nm operate at lower flow rates than larger emitters \cite{yuill2013electrospray, kohigashi2016reduced}. Highly conductive ionic liquids typically have ratios well above this threshold. In the present study $D_e / r_G > 10^3$ for all the liquids, emitter sizes, and flow rates tested. However, as will be shown later, the length of the transition region can become less than two orders of magnitude smaller than the emitter diameter. Although this still represents a significant size disparity, the operation of the electrospray may not be fully decoupled from far-field features.

Finally, self-heating due to dissipation is a key phenomenon in cone-jets of ionic liquids. Ohmic and viscous dissipation are significant in the transition region and continue further downstream along the jet, causing an increase in the liquid temperature and an irreversible voltage drop \cite{gamero2010energy,magnani2024analysis}. Self-heating is significant only for highly conductive liquids such as those investigated in this study. The increase in temperature drastically alters the liquid's physical properties, notably the conductivity, viscosity, and vapor pressure, all of which depend exponentially on temperature. The temperature increase obtained from numerical solutions are well fitted by the power law \cite{magnani2025}:

\begin{equation}
    \Delta T = b_1 \dot{m}^{-b_2} + b_3 
    \label{eq:tempincrease}
\end{equation}
Table \ref{tab:tempincrease} list the values of the fitting parameters for the temperature increase at the point where the surface and conduction currents are equal (approximately the midpoint of the transition region), and at a jet position \(500 \times r_G\) from the vertex of the cone (near the jet break up), for the ionic liquids investigated.

    \begin{figure*}[]\centering 
	\includegraphics[width=\textwidth]{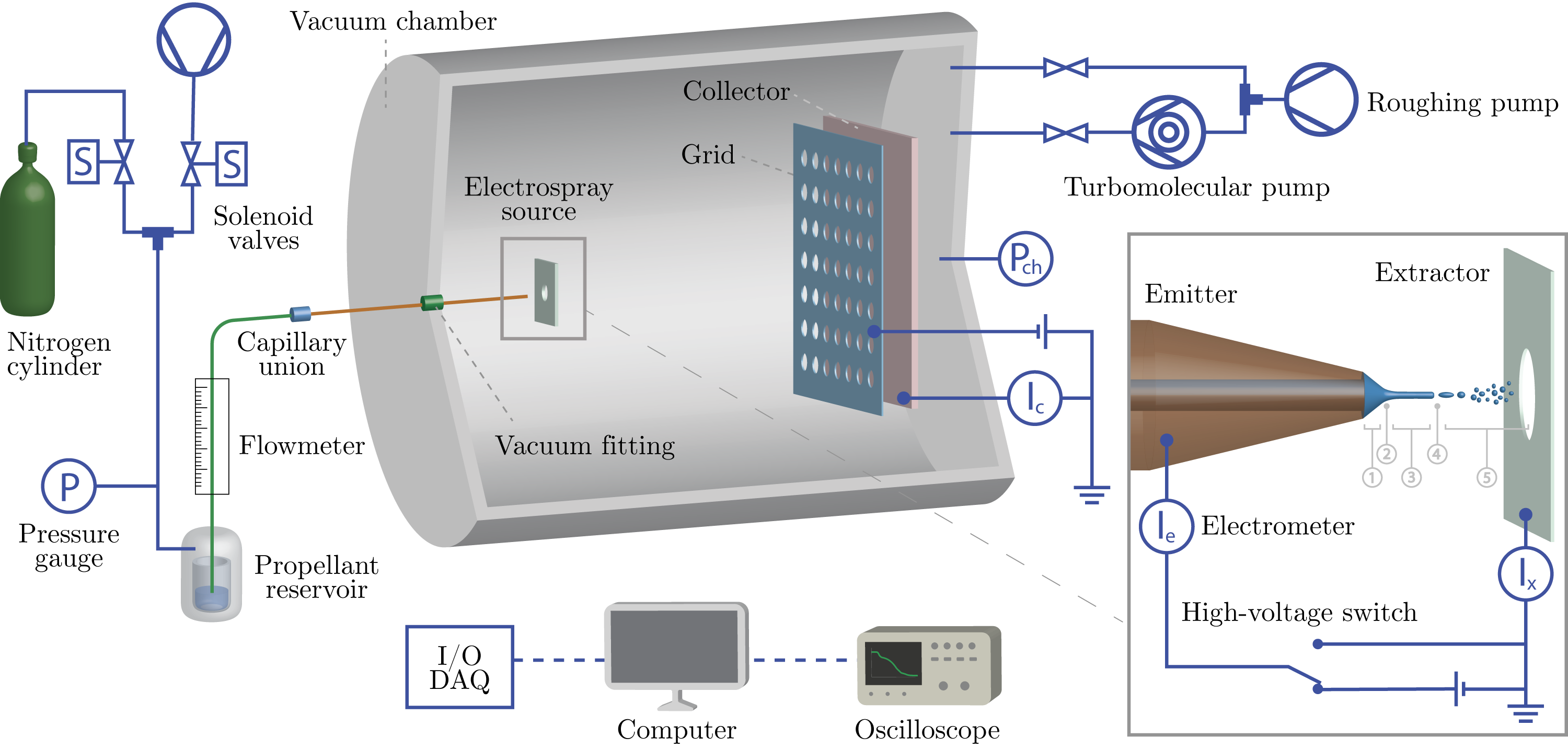}
	\caption{Experimental setup. Cone-jet regions: (1) meniscus, (2) transition region, (3) jet, (4) jet breakup, (5) acceleration region.}
	\label{fig:expsetup}
\end{figure*}

    \section{Methodology}
    \subsection{Experimental setup}

A schematic of the experimental setup is shown in Fig.\ \ref{fig:expsetup}. The electrospray source operates inside a vacuum chamber evacuated to $10^{-3}\,$\SI{}{\pascal} using a turbomolecular pump backed by a roughing pump. The source consists of a \SI{360}{\micro\meter} outer diameter fused silica tube, tapered at a \ang{15} half-angle to form a sharp tip where the outer and inner diameters converge. The tip is sputter-coated with an iridium layer a few tens of nanometers thick, providing electrical connectivity between the liquid and an aluminum disc with a central hole through which the emitter capillary passes. The disc is connected to a high-voltage power supply through a nano-ammeter and a high-voltage switch, the latter used to generate time-of-flight (TOF) signals.

To achieve a manageable hydraulic resistance, the length of the fused silica tube is adjusted from \SI{20}{\centi\meter} for smaller diameter capillaries to \SI{75}{\centi\meter} for larger ones. The emitter capillary exits the vacuum chamber via a vacuum fitting and connects through a zero-dead-volume union to a flow meter capillary that is \SI{50}{\centi\meter} long with an inner diameter ranging from \SIrange{100}{200}{\micro\meter}. Larger diameter flow meter capillaries are used with larger diameter emitters to ensure that the hydraulic resistance of the flow meter is negligible compared to that of the emitter. The end of the flow meter is immersed in a vial containing an ionic liquid, placed inside a larger hermetic bottle. The bottle is filled with nitrogen, and the pressure difference between the bottle and the chamber induces a proportional propellant flow rate. The pressure can be adjusted using a manifold connected to a vacuum pump and a nitrogen cylinder via solenoid valves. Drierite desiccant surrounds the vial to absorb any moisture in the system.

\begin{figure*}[]\centering
	\includegraphics[width=3.25in]{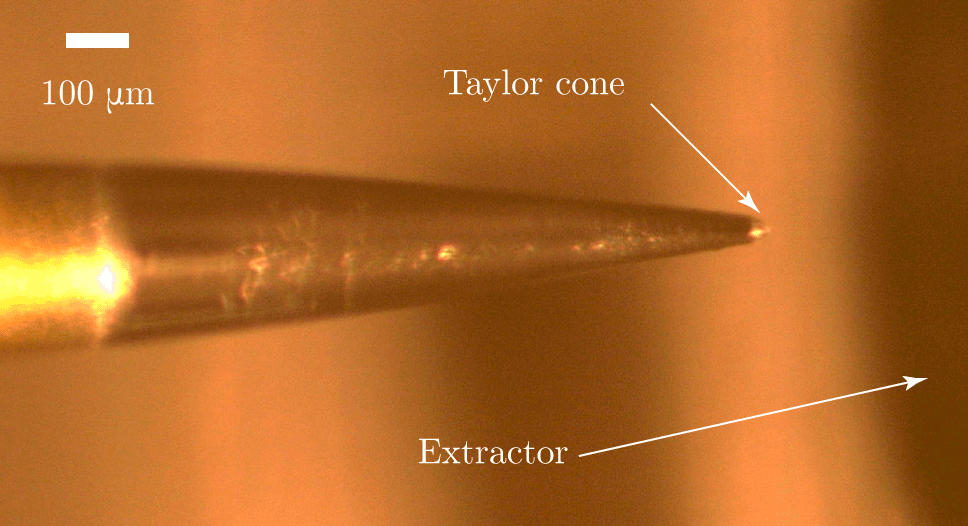}
	\caption{Capillary emitter with a 30\,\textmu m diameter tip.}
	\label{fig:microscope}
\end{figure*}

A grounded extractor electrode faces the emitter at approximately
1 mm from its tip, creating a strong electric field. The emitter voltage is adjusted to form a stable cone-jet electrospray, with the liquid meniscus resembling an ideal Taylor cone. Only positive voltages were used in this study. The beam exits through a \SI{1}{\milli\meter} diameter hole at the center of the extractor. From there, it travels in field-free region until intercepted by a planar collector mounted on a three-axis motorized stage, allowing control of the flight distance $L$. The collector is grounded through an fast-response electrometer connected to an oscilloscope. A stainless steel grid held at \SI{-10}{\volt} is placed before the collector to prevent secondary electrons from escaping. No differences in the emitter current or collector currents were observed for grid voltages ranging from \SI{-4}{\volt} to \SI{-30}{\volt}.

Data acquisition cards (DAQ) are used to record the reservoir pressure $P(t)$, the emitter current $I_e(t)$, the extractor current $I_x(t)$, and the emitter voltage $V(t)$. They also control the pressure at the propellant reservoir via the solenoid valves and adjust the emitter voltage. The collector current $I_c(t)$ is recorded by the oscilloscope when taking a TOF signal, and continuously by the DAQs. The DAQs and oscilloscope are connected to a computer and controlled through a Python script with a graphical user interface. The script automates tasks such as stabilizing and adjusting the reservoir pressure, acquiring time-of-flight (TOF) signals, and post-processing the data. A microscope captures video images of the emitter and meniscus during operation. A photograph of an electrospray using a \SI{30}{\micro\meter} inner diameter capillary is shown in Fig.\ \ref{fig:microscope}. Experiments were conducted at a controlled laboratory temperature of 22$\pm1^\circ$C.

The beam has a slightly positive potential relative to the grounded chamber, increasing near the emitter due to the higher concentration of positive species. To determine whether electrons were being attracted upstream toward the emitter, potentially causing spurious current measurements, we tested an electrospray of BMI-TCM with a large fraction of ions in the beam. We varied the extractor voltage from \SI{0}{\volt} to \SI{-30}{\volt}, while keeping the emitter-extractor voltage difference constant. No differences in the emitter current were observed while varying the extractor potential, indicating that electron attraction toward the emitter is not an issue. Therefore, all experiments were conducted with a grounded extractor.

\subsection{Experimental procedure}
Each ionic liquid was degassed by placing it under vacuum for 24 hours. The mass flow rate was measured directly using a flow meter capillary connected in series with the emitter. When measuring the flow rate, the flow meter line was lifted from the  vial holding the ionic liquid, exposing it to atmospheric pressure and forming a visible air-liquid interface. We then measured the time $\Delta t$ in which the interface traverses a distance $\Delta x$, yielding the hydraulic resistance $R_H$ of the line

\begin{equation}
    R_H = \frac{P_{\text{atm}}}{A_{\text{flow}} \Delta x / \Delta t}
    \label{eq:hyd_res}
\end{equation}

where $P_{\text{atm}}$ is the atmospheric pressure and $A_{\text{flow}}$ is the cross-sectional area of the flow meter capillary. Measuring $R_H$ directly eliminates uncertainties related to the emitter inner diameter and liquid viscosity that would arise if using the Hagen-Poiseuille law. Since the hydraulic resistance of the flow meter capillary is negligible compared to that of the emitter, the position of the interface does not affect the total hydraulic resistance. This procedure was performed for each combination of emitter and liquid, allowing calculation of the flow rate $\dot{m} = \rho P/R_H$ at any pressure $P$. The mass flow rate measured directly using this method is referred to as the total mass flow rate $\dot{m}$. The relationship $\dot{m}(P)$ was verified to be linear and passing through the origin by measuring the interface advancement at different pressures. The pressure head was minimized by maintaining the capillaries and the emitter at the same height. The relative uncertainty of the total mass flow rate was approximately 10\% for all measurements, as discussed in the Supplemental material Section.

When characterizing a liquid the electrospray was initially operated at the maximum reservoir pressure, and from this point the flow rate was gradually decreased until the emission became unstable. For each recorded flow rate within this range, the emission was maintained for at least 5 minutes to assess stability. At high flow rates, the emission was stable for all liquids except for EMI-TFA, which exhibited oscillations in the emitted current. As the flow rate decreased, the emission for all liquids became erratic below a certain threshold, characterized by non-periodic oscillations in the emitted current. Further reduction in flow rate led to intermittent cessation of emission and retraction of the meniscus. An experimental point was considered stable if the emission was sustained for at least 90\% of the test duration and spontaneously recovered after interruptions. We define the minimum flow rate as the lowest flow rate meeting this stability criterion. Prior to taking TOF measurements, the emitter current was sampled at 10\,Hz and averaged over a 2-minute window to obtain the average current $I$, ensuring a low standard error. In each experiment new emitter, flow meter capillary and unions were used to prevent mixing of residual liquids, which may had caused inconsistencies in previous experiments~\cite{caballero2024}.

TOF measurements were performed by periodically shorting the emitter voltage to ground and subsequently reapplying it at a rate of $0.1-1$\,Hz. The resulting $I_c(t)$ current measured at the collector is the TOF signal. A sufficient number of TOF samples (ranging from 10 to 100) were collected for each operating point to ensure low standard errors in performance parameters. Before filling the source with ionic liquid, TOF signals with no emission were recorded to generate a baseline noise signal, which was later subtracted from the TOF data. Each TOF signal was processed by applying a 1\,\textmu s window LOWESS filter and normalized. The normalized signal is the complement of the cumulative probability function of the time-of-flight distribution,  $1-F_\tau(t)$, and the negative of its derivative is the probability density function $f_\tau (t)$. 

The thrust, $I_\text{sp}$, and efficiency were calculated for each TOF signal individually, and we report their mean values and uncertainties. The voltage transition from high voltage to ground occurred over approximately \SI{2}{\micro\second}. The uncertainty in the exact moment of emission cessation was minimized by identifying the time-of-flight of the fastest ions at different extractor-collector distances and extrapolating to zero distance, reducing the uncertainty of the origin of the time variable to \SI{0.2}{\micro\second}. 

\begin{figure*}[]
  \centering
  \includegraphics[width=3.25in]{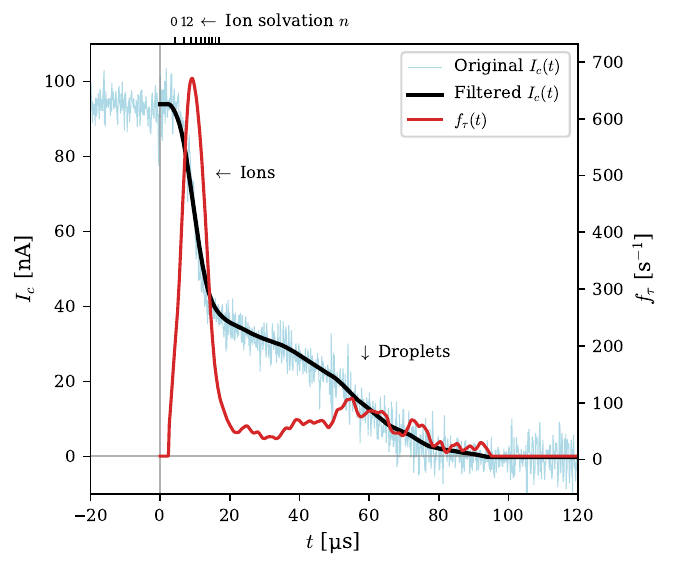}
  \caption{Processing of a typical TOF signal: original collector current signal, collector current signal after LOWESS filter, and probability density function $f_\tau(t)$.} 
  \label{fig:tof_example}
\end{figure*}

\subsection{Performance estimation with time-of-flight}\label{sec:tof}
The integration of the TOF signal provides estimates of the thrust, mass flow rate, specific impulse and propulsive efficiency of the beam, albeit with inherent biases. The mass-to-charge ratio $\xi$ of a particle in the TOF curve is given by

\begin{equation}
    \xi = \left( \frac{t \cos \theta}{L} \right)^2 2 V_r
    \label{eq:xi_map}
\end{equation}

where $t$ is the time-of-flight, $L$ is the axial flight distance, $\theta$ is the angle between the trajectory of the particle and the axis, and $V_r$ is the retarding potential of the particle. $L$ is the distance  between the extractor and collector, and $t$ is obtained directly from the TOF signal. Although our experimental setup does not provide direct measurements of $\theta$ or $V_r$, these variables can be bounded to estimate biases. The differential of beam current carried by particles with mass-to-charge ratios between $\xi$ and $\xi + d\xi$ is given by 

\begin{equation}
    dI_{\text{ch}}  =  I\,f_\Xi(\xi)d\xi 
    \label{eq:diff_Ibeam}
\end{equation}

where $f_\Xi(\xi)$ is the probability density function of the mass-to-charge ratio. The mass flow rate of the charged particles in the beam is then 

\begin{equation}
    \dot{m}_{\text{ch}} = \int \xi \, dI_{\text{ch}} = I \,\int_0^\infty  \xi f_\Xi(\xi) \, d\xi 
    \label{eq:mdotj}
\end{equation}

We simplify the analysis by assuming azimuthal symmetry and that $t$, $V_r$ and $\theta$ in Eq. \eqref{eq:xi_map} are independent, so that $f_\Xi(\xi) =  f_{\tau}(t) f_{V_r}(V_r)\sin \theta f_\Theta(\theta) $:  

\begin{gather}
    \dot{m}_{\text{ch}} \cong \underbrace{\frac{2 V I}{L^2} \left( \int_0^\infty t^2 f_\tau(t) \, dt \right)}_{\dot{m}'} {\left(\mathcal{B}_\theta {\mathcal{B}_c}\right)}^{-1}  \label{eq:mdotj-1}  \\ 
     \mathcal{B}_\theta = \left(\int_{0}^{\theta_{\text{max}}} \sin \theta \cos^2 \theta \, f_\Theta(\theta) \, d\theta \right)^{-1} \\ 
     \mathcal{B}_c = \left( \int_0^\infty \frac{V_r}{V} f_{V_r}(V_r) \, dV_r \right)^{-1}
\end{gather}    

where $V$ is the emitter potential. The term $\mathcal{B}_\theta$ represents the angular bias due to beam spreading, and $\mathcal{B}_c$ is the voltage bias accounting for voltage losses due to energy dissipation in the cone-jet and the formation of surface. Note that the TOF mass flow rate $\dot{m}'$ is obtained from measured quantities. Throughout this text, we use a prime ($'$) to denote quantities estimated with the TOF signal. The TOF mass flow rate is larger than the mass flow rate of charged particles because both $\mathcal{B}_c$ and $\mathcal{B}_\theta$ are necessarily larger than $1$. The ratio $\dot{m}'/\dot{m}$ is thus a biased estimator of the fraction of the beam ejected as charged particles. At high flow rates all the propellant is ejected to the field-free region as charged particles, allowing us to bound the biases, $\mathcal{B}_\theta \mathcal{B}_c \approx \dot{m}'/\dot{m}$ \textit{at high flow rate} ($\dot{m}$ is the mass flow rate fed to the emitter, measured with the capillary flow meter). Note also that if a portion of the mass flow rate is ejected in the form of neutral species, $\dot{m}'$ may be lower than the total mass flow rate $\dot{m}$. Thrust is estimated as

\begin{gather}
    T_{\text{ch}} = \int \frac{L}{t} \, d\dot{m}_{\text{ch}} \cong \underbrace{\frac{2 V I}{L} \left( \int_0^\infty t f_\tau(t) \, dt \right)}_{T'} (\mathcal{B}_\theta \mathcal{B}_c)^{-1}
    \label{eq:thrust_mapped} 
\end{gather}

where the thrust derived from the TOF signal $T'$ exhibits the same biases as $\dot{m}'$. The specific impulse is defined as 

\begin{equation}
    I_{\text{sp}} = \frac{T}{g_0 \dot{m} } \cong \underbrace{ \frac{T'}{g_0 \dot{m}} }_{ I'_{\text{sp}}} (\mathcal{B}_\theta \mathcal{B}_c)^{-1}
    \label{eq:isp}
\end{equation}

The estimate of the specific impulse $I'_{\text{sp}}$ is computed with the total mass flow rate (measured with the capillary flow meter) and the thrust obtained from the TOF curve. This estimate has the same biases as $\dot{m}'$ and $T'$. For each liquid, emitter size and flow rate, different values of the emitter voltage $V$ were required to stabilize the electrospray. These variations are eliminated by plotting the thrust and $I_\text{sp}$  at a common acceleration voltage. Throughout the article we use $V_a = \SI{10}{\kilo\volt}$ as this common value. Since thrust is proportional to the effective exhaust velocity, with $T = \dot{m} c$ and $c \propto \sqrt{V_a}$, we scale the thrust and specific impulse as:

\begin{equation}
    T'_a = T' \sqrt{\frac{V_a}{V}}, \qquad   
     I'_{\text{sp,}a} = I'_{\text{sp}} \sqrt{\frac{V_a}{V}}\label{eq:10kv}
\end{equation}

Although an acceleration voltage of 10\,kV is higher than in typical plasma thrusters, acceleration voltages as high as \SI{30}{\kilo\volt} are currently used in commercial FEEP thrusters~\cite{buldrini2023results}. The propulsive efficiency is defined as

\begin{equation}
    \eta = \frac{ T^2 }{2 \dot{m} P_{\text{in}}}   \cong \frac{\dot{m}_{\text{ch}}}{\dot{m}} \underbrace{\frac{ T'^2 }{2 \dot{m}' IV}}_{\eta'} (\mathcal{B}_\theta \mathcal{B}_c)^{-1}
\end{equation}
where $P_{\text{in}} = IV$ is the input power. We assume that the thrust produced by neutral particles, if they exist in the beam, is negligible. In addition to the biases $\mathcal{B}_\theta$ and $\mathcal{B}_c$, the efficiency $\eta'$ obtained from the TOF curve may further overestimate the propulsive efficiency if a portion of the propellant mass is emitted as neutrals ($\dot{m}_{\text{ch}}/\dot{m}<1$). 

Figure \ref{fig:tof_example} illustrates the processing of a TOF signal for BMI-TCM, electrosprayed from a \SI{15}{\micro\meter} emitter  with a flow rate $\dot{m} = 2.21 \times 10^{-11}$\,kg/s. The  current of the electrospray was \SI{309}{\nano\ampere}, and the emitter voltage was set to \SI{1098}{\volt}. We also show the probability density function to better visualize the distribution of ions  and droplets. The values of the thrust and specific impulse at \SI{10}{\kilo\volt} were $T'_a = \SI{311}{\nano\newton}$ and $I'_{\text{sp,}a} = \SI{1573}{\second}$, while the efficiency and the ratio between TOF and total mass flow rate were $\eta' = 0.707$ and $\dot{m}'/\dot{m} = 1.097$. 

\section{Results}
\begin{figure*}[t!]
  \centering
  \includegraphics[width=\textwidth]{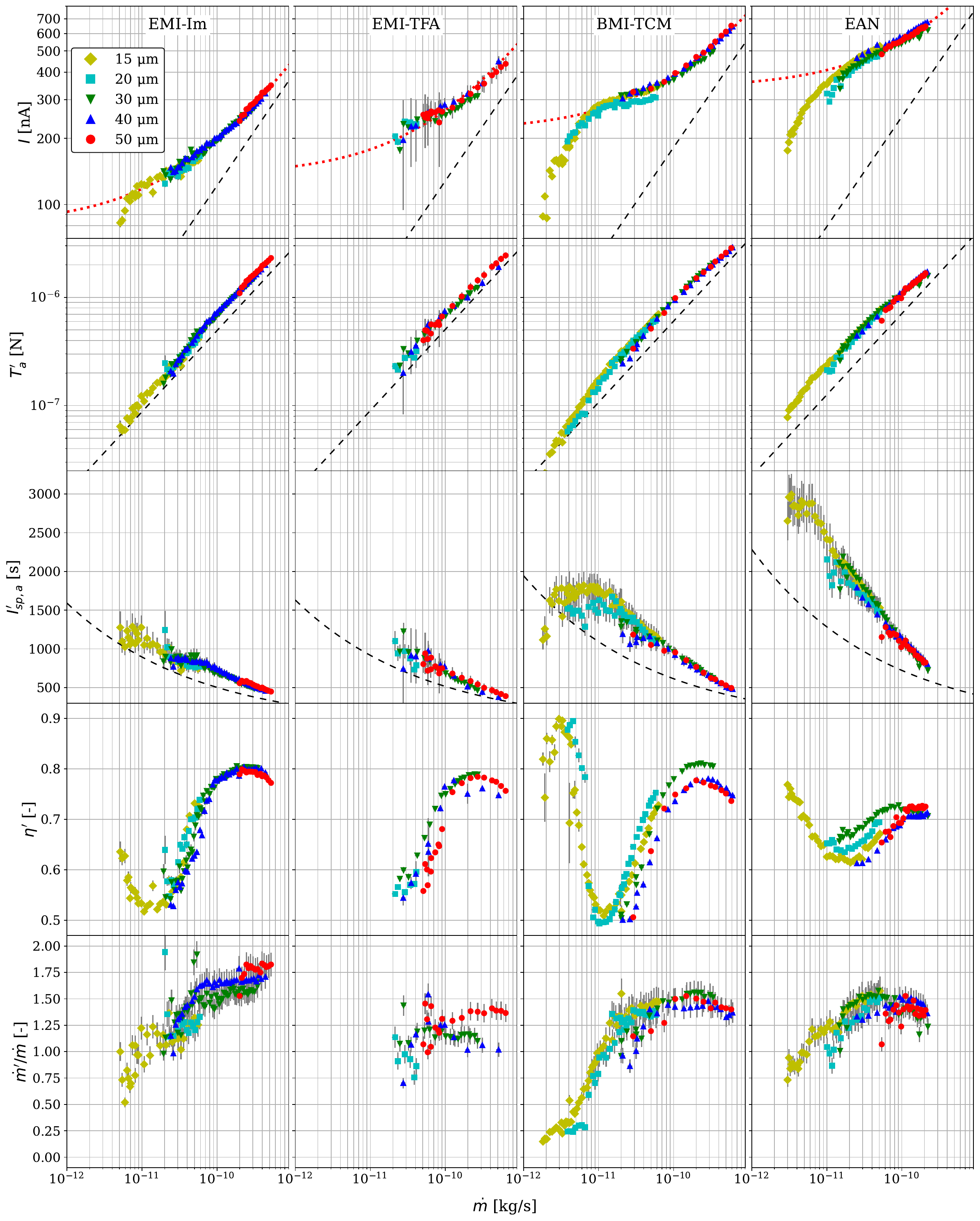}
  \caption{Emitter current, thrust and  specific impulse at 10\,kV, efficiency  and ratio $\dot{m}'/\dot{m}$ as functions of the total mass flow rate.}
  \label{fig:results}
\end{figure*} 

Figure \ref{fig:results} shows various performance parameters as functions of the total mass flow rate, for each liquid and emitter diameter. These parameters include the emitter current, the thrust and specific impulse at an acceleration voltage of \SI{10}{\kilo\volt} ($T'_a$ and $I'_{\text{sp,}a}$), the efficiency ($\eta')$, and the ratio between the TOF-derived mass flow rate and the total mass flow rate ($\dot{m}'/\dot{m}$). Measurement uncertainties are represented by vertical and horizontal lines, and their derivations are detailed in the Supplemental material section. The uncertainty bars do not include the biases described in Section \ref{sec:tof}. In these experiments the emitter voltage was set such that the meniscus had a conical shape similar to an ideal Taylor cone. This value varied from approximately \SIrange{700}{1500}{\volt}, depending on the liquid, flow rate, and capillary size. Higher flow rates and emitter diameters required higher voltages. At higher flow rates, the cone-jet was stable within a range of about \SI{100}{\volt}. The stability range narrowed with decreasing flow rate, being only \SIrange{5}{10}{\volt} at the minimum flow rates, consistent with the known stability envelope of cone-jets \cite{cloupeau1989electrostatic}. No appreciable changes in the emitter current or beam composition were observed when the emitter voltage was varied within the stability range at a fixed flow rate. Note that when the electrospray is stable, the value of any performance parameter at fixed mass flow rate is largely independent of the emitter diameter. The black dashed lines in the charts for the emitter current, thrust and specific impulse  correspond to the values associated with the scaling law \eqref{eq:curr_scaling}. Particles with uniform mass-to-charge ratio $\xi = \dot{m}/I$, accelerated by a voltage $V_a=10\,$kV, attain a speed $c = \sqrt{2{V_a}I/\dot{m}}$. Combining this with $T=\dot{m}c$,  $I_\text{sp}=c/g_0$, and Eq. \eqref{eq:curr_scaling} yield the following scaling laws for the thrust and specific impulse:

 \begin{equation}
        T \cong \dot{m}^{3/4} (5.2 V_a)^{1/2} \left(\frac{\gamma K}{\rho}\right)^{1/4}
        \label{eq:thrust_scaling}
    \end{equation}

    \begin{equation}
        I_{\text{sp}} \cong \dot{m}^{-1/4} (5.2 V_a)^{1/2} \left(\frac{\gamma K}{\rho}\right)^{1/4} g_0^{-1}
        \label{eq:isp_scaling}
    \end{equation}
    
Note that Eq. \eqref{eq:curr_scaling} underpredicts the current for all liquids, especially BMI-TCM and EAN. Adding a $y-$intercept 
  
    \begin{equation}
    I \cong I_0 + \psi \left( \frac{ \gamma K\dot{m}}{\rho} \right)^{1/2}
    \label{eq:curr_scaling_offset}
\end{equation}

produces a much better fit of the experimental data; this trend is plotted as red dotted lines. For reference, the fitting parameters $\{\psi,I_0\}$ for EMI-Im, EMI-TFA, BMI-TCM and EAN are  $\{2.48,81 \text{ nA}\}$, $\{2.70,136\text{ nA}\}$, $\{2.41,217\text{ nA}\}$ and $\{2.26,340\text{ nA}\}$ respectively. At $\dot{m} \lesssim 10^{-11}$\,kg/s, the measured current becomes lower than this trend. The failure of the traditional current law for cone-jets, Eq. \eqref{eq:curr_scaling}, is due to dissipation and self-heating of the liquid \cite{gamero2010energy,gamero2019dissipation}. This raises the temperature and therefore the electrical conductivity of the propellant, and ultimately leads to a higher emitted current at fixed propellant flow rate \cite{magnani2024analysis,magnani2025}. Furthermore, the physical properties of the propellant vary along the cone-jet. The intensity of self-heating and the departure from isothermal behavior increases with electrical conductivity, and are of critical importance for the propellants of interest to electrospray propulsion, which must have high conductivities to maximize specific impulse. Another important feature of Fig.~\ref{fig:results} is the significant decrease of the ratio $\dot{m}'/\dot{m}$ when the total mass flow rate falls below a propellant-dependent threshold: approximately $5\times10^{-11}$\,kg/s for EMI-Im, and $2\times10^{-11}$\,kg/s for BMI-TCM and EAN. At higher flow rates, $\dot{m}'/\dot{m}$ remains approximately constant, with the TOF estimate being larger than the value obtained with the flow meter. This is explained by the voltage and angular biases described in Eq.~\eqref{eq:mdotj-1} and will be discussed in Section~\ref{sec:bias}. The sharp decrease of $\dot{m}'/\dot{m}$ at the smallest flow rates indicates that a significant amount of the propellant is ejected in the form of uncharged particles, representing a propellant loss.

\begin{figure}[]
  \centering
  \includegraphics[width=\textwidth]{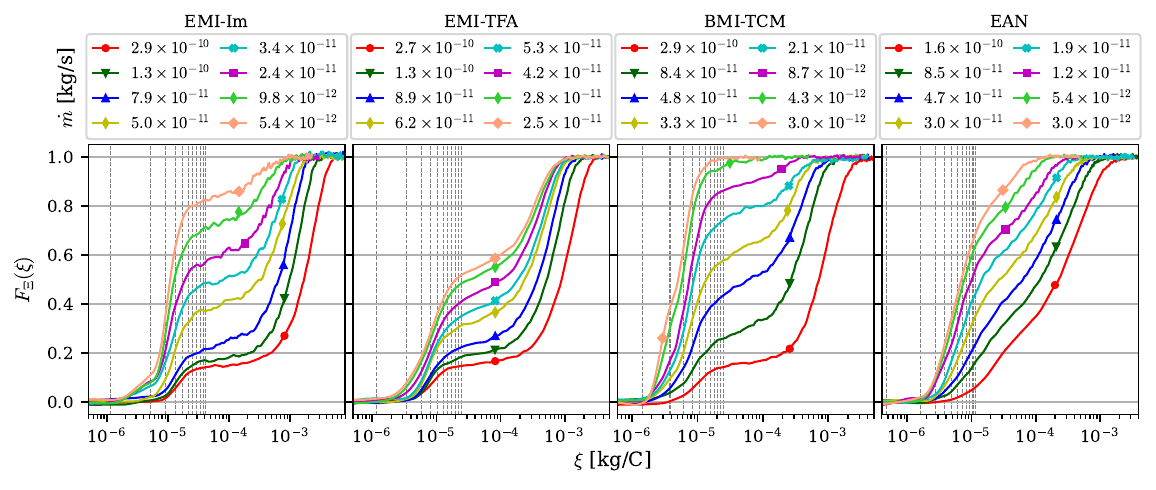}
  \caption{Cumulative distribution functions $F_\Xi(\xi)$ of the mass-to-charge ratio $\xi$ for each liquid and several mass flow rates $\dot{m}$.}
  \label{fig:tofplot}
\end{figure}

Figure~\ref{fig:tofplot} shows the cumulative mass-to-charge ratio distribution obtained by mapping $F_\tau(t) \rightarrow F_\Xi(\xi)$ with Eq.~\eqref{eq:xi_map} and using $\theta = 0$ and $V_r = V$. The vertical dashed lines correspond to the masses of ion clusters with varying solvation states $n$ ($[\text{AB}]_n\text{A}^+$, where A is the cation, B is the anion and $n = 0, 1, \ldots, 10$). There is a clear separation between the ion clusters and droplets for EMI-Im, EMI-TFA and BMI-TCM. This distinction is less evident for EAN because the droplets have smaller mass-to-charge ratios, while the ion clusters have higher solvation states. According to Born's model, the energy barrier $\Delta G$ impeding the evaporation of an ion cluster of radius $a_c$ and charge $q_c$ is given by \cite{thomson1979field, labowsky2000continuum}

\begin{equation}
    \Delta G = 4 \pi \gamma a_c^2 + \frac{q_c^2}{8\pi \varepsilon_0 a_c}\left(1 -\frac{1}{\varepsilon}\right)
    \label{eq:dg0_1}
\end{equation}

\begin{figure}[]
  \centering
  \includegraphics[width=\textwidth]{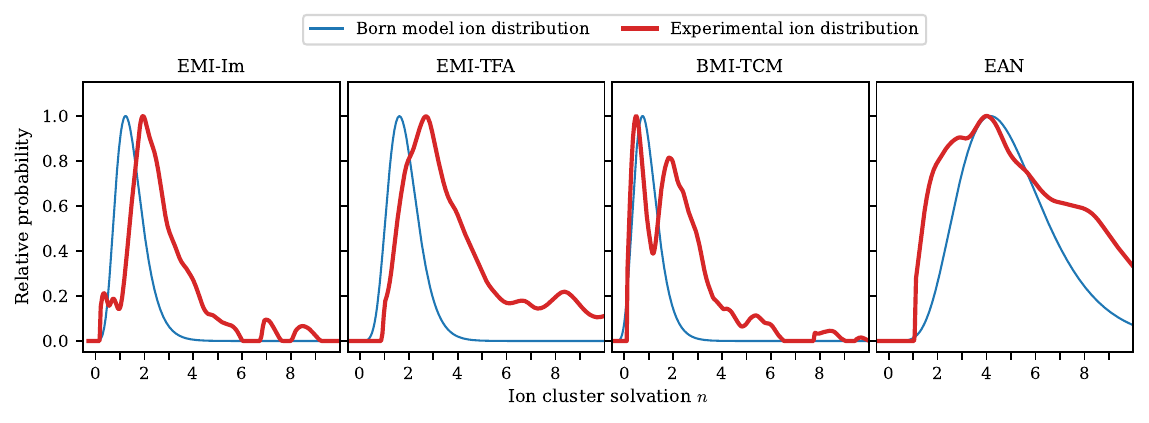}
  \caption{Probability density functions of singly-charged ions obtained from TOF curves and from Born's model. Physical properties are evaluated a the temperature of the jet.}
  \label{fig:ion_distributions}
\end{figure}

The probability for an ion cluster to evaporate is proportional to the Boltzmann factor, $\exp(-\Delta G / k_B T)$ \cite{iribarne1976evaporation}. Assuming singly charged ions and typical values of $\gamma$ and $\varepsilon$, the energy barrier $\Delta G$ is minimized at an ion cluster size of approximately $0.5$\,nm. For the same cluster radius, EAN clusters will contain more molecules because the volume of a cluster with solvation state $n$ can be approximated as $v_c = \frac{4}{3}\pi a_c^3 \approx [(n+1) M^+ + nM^-]/\rho$, and the molecular mass of EAN is much smaller than those of the other liquids. Consequently, the distribution of solvation states is wider for EAN. This is illustrated in Fig.~\ref{fig:ion_distributions}, where the experimental probability density function of the ions' solvation state is compared with that derived from the Born model.

\begin{figure}[]
  \centering
  \includegraphics[width=3.25in]{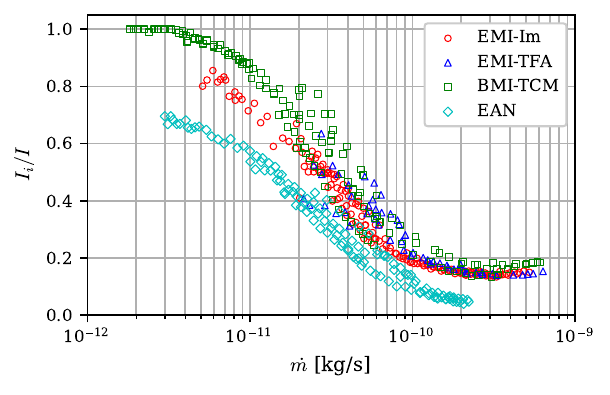}
  \caption{Ratio between the ion current $I_i$ and the total current $I$ of the beam for each liquid (all emitter sizes) as a function of the total mass flow rate}
  \label{fig:ionratio}
\end{figure}

Figure~\ref{fig:ionratio} plots the ratio between the ion current and the total current as a function of the total mass flow rate. Any particle with a mass-to-charge ratio smaller than that of $[\text{AB}]_{10}\text{A}^+$ is considered an ion. Three distinct regimes can be identified: 1) for $\dot{m} \gtrsim 10^{-10}\,\text{kg/s}$, the beams are dominated by droplets, and the ion current ratio is approximately constant (between 0.05--0.1 for EAN and approximately 0.18 for all other liquids); 2) at lower mass flow rates, $4 \times 10^{-12}\,\text{kg/s} < \dot{m} < 10^{-10}\,\text{kg/s}$, there is a droplet–ion mixed regime in which the ion fraction increases as the flow rate decreases; and 3) in the case of BMI-TCM at the lowest flow rates, $\dot{m} \lesssim 4 \times 10^{-12}\,\text{kg/s}$, the beam consists solely of ions. This pure ion regime is typically observed in porous and externally wetted emitters. For instance, 92\% of the current corresponds to ions with solvation numbers $n \leq 10$ in a porous emitter thruster using EMI-BF$_4$~\cite{corrado2024direct}. This regime could not be consistently reached with the other liquids, although we observed it in EMI-Im in earlier experiments with a \SI{20}{\micro\meter} emitter~\cite{caballero2024}. Based on time-of-flight and retarding potential measurements~\cite{gamero2021electrosprays}, ion emission in the droplet-dominated regime originates from either a few droplets or jet pinching zones at the jet breakup. In the droplet–ion mixed regime, ion emission occurs from an increasing number of droplets in the breakup and also from the surface of the jet at the lowest flow rates \cite{Gamero2002}. In the pure ion regime, ion emission takes place from the tip of the cone.

Hysteresis in the current versus mass flow rate behavior was observed at the lowest flow rates for all liquids. The minimum flow rate could only be accessed by operating the electrospray at high flow rates and voltages and gradually lowering them. Conversely, increasing the flow rate starting from a low unstable value could only stabilize the cone-jet above the minimum flow rate. EMI-TFA electrosprays were much more challenging to stabilize, exhibiting erratic emission behavior despite having physical properties similar to EMI-Im. This suggests that emission stability may depend not only on bulk physical properties and emitter geometry but also on the molecular properties of the liquid. This led to higher variance in the EMI-TFA measurements, as the current and beam composition changed considerably between measurements at fixed flow rate. We did not electrospray EMI-TFA with the \SI{15}{\micro\meter} emitter due to difficulties encountered with the \SI{20}{\micro\meter} emitter.

The time-of-flight distance for all reported data was $L = 0.1648\,\text{m}$. If the collector was moved further away from the electrospray source, a decreasing ion current was observed in the TOF signal at sufficiently large separations, indicating that the outer regions of the beams, which preferentially carry ions, missed the collector. This leads to erroneously larger TOF thrust and mass flow rate calculations. We verified that the entire beam was captured by the collector when $L = 0.1648\,\text{m}$, as the collector current was approximately equal to the emitter current times the optical transparency of the grid, and there were no differences in the collected current when shortening $L$. In contrast, when operating BMI-TCM in the pure ionic regime, no appreciable differences in the TOF signal were observed at varying collector distances.

The BMI-TCM propellant in the vial experienced significant darkening after a few days under vacuum. The cause of the color change is unknown to us. No cyanide was detected with a Honeywell BW Solo HCN sensor upon exposure of the propellant reservoir and the vacuum chamber to atmospheric pressure. In the earliest phase of our research~\cite{caballero2024}, we discovered that operation at high emitter potentials (\SIrange{2000}{2500}{\volt}) may damage the iridium coating of the emitters, rendering them unusable after a short period of operation.

    \section{Discussion}
    \subsection{Minimum flow rate and emitter size}

\begin{figure}[]
  \centering
  \includegraphics[width=3.25in]{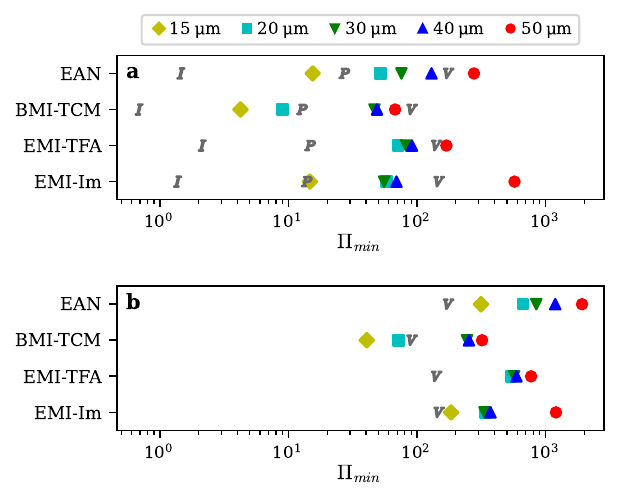}
  \caption{Dimensionless minimum flow rate and minimum flow rate criteria, \eqref{eq:minflow_visc}$-V$, \eqref{eq:minflow_pol}$-P$, \eqref{eq:minflow_iontransport}$-I$: a) $\Pi_\text{min}$ and criteria evaluated at \SI{25}{\degreeCelsius}; b)  $\Pi_\text{min}$ evaluated at current crossover and \eqref{eq:minflow_visc} evaluated at \SI{25}{\degreeCelsius}.  }
  \label{fig:minflowrate}
\end{figure}

The reduction of the minimum stable flow rate with decreasing emitter diameter is a critical result due to its importance for maximizing the specific impulse, Eq. \eqref{eq:isp_scaling}. For example, the minimum flow rates of EAN cone-jets were $5.0 \times 10^{-11}$\,kg/s, $2.3 \times 10^{-11}$\,kg/s, $1.4 \times 10^{-11}$\,kg/s, $9.3 \times 10^{-12}$\,kg/s, and $2.8 \times 10^{-12}$\,kg/s for emitter diameters of  \SI{50}{\micro\meter}, \SI{40}{\micro\meter}, \SI{30}{\micro\meter}, \SI{20}{\micro\meter}, and \SI{15}{\micro\meter} respectively. In some cases, decreasing the emitter diameter did not result in a significantly lower minimum flow rate, such as for \SI{20}{\micro\meter} versus \SI{40}{\micro\meter} emitters in the EMI-Im experiments; however, the minimum flow rate was never higher for a smaller emitter. The substantial decrease in the minimum flow rate of more than one order of magnitude resulted in roughly doubling the $I_\text{sp}$ for all liquids except EMI-TFA. Figure \ref{fig:minflowrate} shows the dimensionless minimum flow rate for each liquid and emitter diameter, along with criteria \eqref{eq:minflow_visc}, \eqref{eq:minflow_pol}, and \eqref{eq:minflow_iontransport}. Full charge separation, Eq. \eqref{eq:minflow_iontransport}, is evaluated with the minimum experimental current and assuming an ionicity $\nu = 0.7$ for all liquids. The minimum flow rate was always above the charge separation limit for all liquid/emitter combinations. The polarization limit \eqref{eq:minflow_pol} is near the minimum flow rate in some cases, however the product $\varepsilon Re_K$ ranges between 0.1 and 0.15 and therefore this limit is not applicable. The viscous limit matches the minimum flow rates for the larger emitter diameters but overpredicts them for the smaller ones. However, it must be noted that the values of both the dimensionless flow rates and the viscous criterion \eqref{eq:minflow_visc} in Fig. \ref{fig:minflowrate}.a are hindered by the strong self-heating  exhibited by these cone-jets. The higher and spatially varying temperature is not considered in Fig. \ref{fig:minflowrate}.a, where all physical properties are evaluated at 25\textdegree{}C. For example, at the minimum flow rates measured for EMI-Im and EAN and for an upstream temperature of 25\textdegree{}C, the temperatures at various locations differ significantly. The temperature at the base of the jet (the axial position where the curvature is maximum) is 31\textdegree{}C for EMI-Im and 34\textdegree{}C for EAN. At the current crossover point, the temperatures are 460\textdegree{}C and 556\textdegree{}C, respectively. Further downstream, at a distance $500\times r_G$ from the vertex of the cone, the temperatures rise to 555\textdegree{}C and 711\textdegree{}C \cite{magnani2025}. These large temperature increases are especially important in the evaluation of the electrical conductivity and viscosity, which depend exponentially on temperature. Obviating axial temperature variations, a dimensionless flow rate calculated with the electrical conductivity evaluated at the temperature of the current crossover represents the state of the cone-jet better than if it is evaluated at 25\textdegree C. On the other hand, the best estimate of the viscous criterion \eqref{eq:minflow_visc} results from an evaluation at the temperature in the base of the jet, which is where viscous dissipation is dominant. Figure \ref{fig:minflowrate}.b shows these better estimates, indicating a better agreement of the viscous criterion. However, the results are inconclusive and suggest that, given the large variation of temperature, formulating a simple minimum flow rate criterion for ionic liquids may not be possible.      

\begin{figure}[]
  \centering
  \includegraphics[width=3.25in]{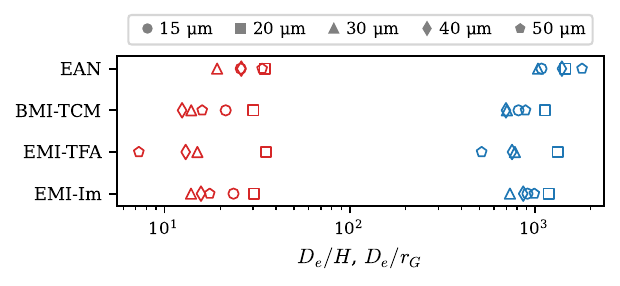}
  \caption{Ratios $D_e/H$ (red markers) and $D_e/r_G$ (blue markers) evaluated at the minimum flow rate for each liquid and emitter.}
  \label{fig:transition}
\end{figure}

Figure \ref{fig:transition} presents the ratios $D_e/r_G$ and $D_e/H$ between the emitter diameter $D_e$, the characteristic length of the transition region $H$, and the characteristic jet diameter $r_G$, evluated at the minimum flow rates. While $D_e/r_G$ exceeded $10^3$ in all cases, $D_e/H$ is between 30 and 230. Although $D_e/H$ is relatively large, it may not be large enough to eliminate the geometry of the emitter as a factor determining the minimum flow rate and its dependence on the diameter of the emitter. Geometrical features of the electrodes may influence
the local distribution of charge in the cone-jet, potentially affecting its stability at low
flow rates. 

The observation that smaller menisci are more stable is likely applicable to externally wetted and porous emitter electrosprays as well. In a study using externally wetted emitters with radii of curvature ranging from \SIrange{2.5}{60}{\micro\meter}, it was observed that smaller tip radii allowed emission at lower currents, thereby forming stable electrosprays at lower flow rates \cite{castro2009effect}. The influence of emitter diameter on the minimum flow rate could also explain difficulties in reproducing certain experimental studies. For instance, the minimum flow rate of a mixture of EAN and sulfolane showed discrepancies of an order of magnitude between studies \cite{alonso2014search, caballero2022effect}. The emitters used in \cite{alonso2014search} were made by pulling fused silica capillaries under a torch, resulting in tips of about \SI{15}{\micro\meter} to \SI{20}{\micro\meter} in diameter, whereas \cite{caballero2022effect} employed emitters mechanically tapered to \SI{100}{\micro\meter} in diameter at the tip.

We did not observe a dependence of the minimum flow rate on the hydraulic resistance of the emitter. For example, the hydraulic resistances of the \SI{15}{\micro\meter} emitters were very similar to those of the \SI{20}{\micro\meter} emitters, as shorter tube lengths were employed for the \SI{15}{\micro\meter} emitters to operate them at manageable reservoir pressures. Nonetheless, the \SI{15}{\micro\meter} emitters had a considerably lower minimum flow rate than the \SI{20}{\micro\meter} emitters. An additional test compared the minimum flow rate of \SI{50}{\micro\meter} emitters with a 5-fold difference in hydraulic resistances. This test, carried out with BMI-TCM, did not yield appreciable differences in the minimum flow rate.

\subsection{Comparison between the propulsive characteristics of EMI-Im, EMI-TFA, BMI-TCM and EAN}
The specific impulse and the efficiency are key characteristics determining the quality of a propellant. Figure \ref{fig:results} shows that for a given mass flow rate, EAN yields the highest $I_\text{sp}$, followed by BMI-TCM, EMI-TFA and EMI-Im in this order. An approximate ranking can be made \textit{a priori} based solely on the values of their physical properties: Eq. \eqref{eq:isp_scaling} shows that high electrical conductivity, high surface tension, and low density favor $I_\text{sp}$. In fact, EAN, BMI-TCM, EMI-TFA and EMI-Im are ranked in this order according to their $\gamma K/\rho$ ratios. This simple criterion is only qualitative due to several factors: self-heating drastically alters the relationship between electrospray current and mass flow rate; the minimum flow rate, which is  unknown \textit{a priori}, determines the minimum mass-to-charge ratio of the droplets; and propellant losses in the form of ejected uncharged particles affect the $I_\text{sp}$ at sufficiently low flow rates. Regardless of this, Fig. \ref{fig:results} shows that both EAN and BMI-TCM can provide specific impulses sufficiently high for most propulsion needs.

The efficiency curves in Fig. \ref{fig:results} are also interesting. The efficiency of BMI-TCM, EMI-TFA, and EMI-Im is maximal at high flow rate (approximately 80\%), decreases rapidly for $\dot{m} \lesssim 10^{-10}$\,kg/s, and in the case of BMI-TCM and EMI-Im, increases again for $\dot{m} \lesssim 10^{-11}$\,kg/s. The concept of polydispersive efficiency, i.e. the reduction of efficiency due to the presence of particles with very different charge-to-mass ratios and therefore velocities, together with figures \ref{fig:tofplot} and \ref{fig:ionratio} explain these trends: the current fraction of droplets is relatively constant at high flow rates, approximately 0.82, and the mass-to-charge ratios of droplets and ions differ by over two orders of magnitude. Thus, the efficiency is bounded by the current fraction of droplets (the population of ions, although consuming power proportionally to its current, contributes a negligible amount of thrust due to the very low mass-to-charge ratios and therefore low mass flow rates). At lower flow rates, $\dot{m} \lesssim 10^{-10}$\,kg/s, the current fraction of droplets decreases monotonically, and the efficiency decreases in the same proportion bounded by the droplet current fraction. At very low flow rates, the efficiency of BMI-TCM increases because the electrospray operates in the pure ion mode (charged droplets disappear from the beam), whereas in the case of EMI-Im, the mass-to-charge ratios of ions and droplets become relatively similar. The efficiency curve for EAN has a similar shape, but the minimum efficiency at intermediate flow rates is higher than for BMI-TCM, EMI-TFA, and EMI-Im. The efficiency of EAN ranges between 61\% and 77\%. This is due to the very high electrical conductivity of EAN, which produces droplets with very low mass-to-charge ratios, not far from those of ions (see Fig. \ref{fig:tofplot}).    

\subsection{Thrust bias in the time-of-flight estimation} \label{sec:bias}

Propellant is ejected as uncharged particles at the lowest flow rates, likely due to the evaporation of neutrals caused by strong self-heating. Conversely, at high flow rates, reduced self-heating minimizes these mass losses. When the mass losses are negligible, the ratio $\dot{m}'/\dot{m}$ approximates the bias $\mathcal{B}$ in the thrust measurement derived from the TOF curve: 

\begin{equation}
   \mathcal{B} = \mathcal{B}_\theta\, \mathcal{B}_c = \frac{T'}{T} \cong \left. \frac{\dot{m}'}{\dot{m}} \right|_{\dot{m} \gtrsim 5\times10^{-11}\,\text{kg/s}}
\end{equation}

We expect a slight decrease of $\mathcal{B}$ with decreasing mass flow rate, as both energy dissipation decreases~\cite{gamero2021electrosprays} and the beam narrows~\cite{gamero2008structure, thuppul2021mass} with decreasing mass flow rate. Given the uncertainty in the TOF measurement, it is not possible to observe this trend in the $\dot{m}'/\dot{m}$ data presented in Fig.~\ref{fig:results}. Therefore, we use the average of $\dot{m}'/\dot{m}$ for $\dot{m} \gtrsim 5\times10^{-11}$\,kg/s to estimate an upper bound for $\mathcal{B}$. The maximum values of $\mathcal{B}$ for EMI-Im, EMI-TFA, BMI-TCM, and EAN are 1.75, 1.4, 1.5, and 1.5, respectively.

The voltage and angular biases can be estimated using data from previous studies. The voltage loss can be calculated as the voltage applied to the emitter minus the kinetic energy per unit charge of the jet at the breakup location. This quantity has been measured for cone-jets of EMI-Im down to flow rates in the upper range considered in this study~\cite{gamero2021electrosprays}. For example, for a mass flow rate of $2.3 \times 10^{-10}$\,kg/s, the voltage loss is \SI{151}{\volt}. Using this voltage loss and considering that we employed an emitter voltage of approximately \SI{1250}{\volt} for our EMI-Im cone-jets at high flow rates, we estimate $\mathcal{B}_c = 1.14$ as a lower bound for the voltage bias in our cone-jets.

The angular bias depends on the radial expansion of the beam. This expansion is determined by both space charge effects and the electric field in the emitter-extractor region, which in turn depend on the geometry of the electrodes and the applied voltage difference. The current density in electrosprays is approximately constant for low polar angles and decreases rapidly above a certain threshold polar angle. A simple approach is to model the angular distribution using a constant current density, $f_\Theta(\theta) = constant$, for angles below the cone half-angle, $\theta < \theta_{1/2}$, and zero for larger angles. In that case, the angular bias is:

\begin{equation}
    \mathcal{B}_\theta = \left[ \frac{1}{3} \left( 1 + \cos\theta_{1/2} + \cos^2 \theta_{1/2} \right) \right]^{-1}
\end{equation}

Alternatively, the angular distribution can be fitted more accurately to a generalized Gaussian with zero mean, which exhibit near-constant densities at low polar angles but has smooth tails. Defining a cone width $\theta_{w}$ such that $f_\Theta(\theta_{w}) = \frac{1}{2} f_\Theta(0)$ and a sharpness parameter $S$, the distribution is given by

\begin{equation}
    f_\Theta(\theta) \propto \exp \left[ -\left( \frac{\theta}{\theta_{w}} \right)^S \ln 2 \right]
\end{equation}

Fitting data from previous studies for EMI-Im provide parameter values within the ranges $\theta_{w} = 15^\circ$--$20^\circ$ and $S = 2.5$--$6.5$~\cite{gamero2013expansion, thuppul2021mass}, for which $\mathcal{B}_\theta = 1.04$--1.14. However, these studies were conducted with a different electrode geometry and at higher voltages than those used in the present study. Combined with our estimate $\mathcal{B}_c = 1.14$, this yields a thrust bias between $\mathcal{B} = 1.2$--$1.3$. The maximum values of $\mathcal{B}$ obtained from the ratios $\dot{m}'/\dot{m}$ are slightly larger than expected but are consistent with typical values of the voltage and angular biases. We believe that the geometry of our electrospray source is producing very broad beams, and the resulting high angular bias is the main contributor to $\mathcal{B}$. First, the voltage difference between the emitter and extractor is relatively low, which accentuates the effect of beam expansion in the initial, space-charge-dominated region. Second, the extractor orifice is relatively close to the emitter tip, and the emitter is chamfered at a small angle (see Fig.~\ref{fig:microscope}). As a result, compared to the electrospray source in reference~\cite{gamero2013expansion}, a larger fraction of the emitter-to-extractor potential in our electrospray source is used to increase the radial velocity component of beam particles. 

\section{Conclusions}

This study characterized electrosprays of the ionic liquids EMI-Im, EMI-TFA, BMI-TCM and EAN using capillary emitters with tip diameters from 15\,\textmu m to 50\,\textmu m. The thrust was measured indirectly using the time-of-flight technique, and the mass flow rate was determined both directly via a flow meter and indirectly through TOF. Electrosprays from smaller emitters exhibited substantially lower minimum flow rates than those from larger emitters, leading to a notable increase in the maximum attainable $I_\text{sp}$. For example, the maximum $I_\text{sp}$ for EMI-Im increased from 600\,s with the \SI{50}{\micro\meter} emitter to over 1300\,s with smaller emitters when referred to an acceleration voltage of \SI{10}{\kilo\volt}. Using the smallest emitters, specific impulses up to 3000\,s are attainable for EAN. For the same operating flow rate, the thrust, efficiency, and $I_\text{sp}$ remained consistent across different emitter sizes.

The causes for the influence of emitter size on the stability of these electrosprays remain unclear. Theories suggest that the minimum flow rate of a cone-jet is determined by instabilities local to the transition region and should be unaffected by the electrode geometry, provided its characteristic lengths are significantly larger than those of the jet and transition region. In our study, the jets produced were approximately 10\,nm in diameter, many orders of magnitude smaller than the emitter tip diameter, the distance between the emitter and extractor, and the extractor hole. The length of the transition region was comparatively larger than the jet diameter but still about two orders of magnitude smaller than the bases of the Taylor cones formed at the minimum flow rate. The appropriate scaling law for the minimum flow rate for our liquids is Eq.~\eqref{eq:minflow_visc}. Due to significant self-heating in cone-jets of highly conductive liquids, the physical properties vary drastically throughout the cone-jet. As a result, there is a large uncertainty in the evaluation of Eq. \eqref{eq:minflow_visc}, hindering both the verification of this criterion and an \textit{a priori} estimation of the minimum flow rate.

The efficiency inferred from TOF ranged between 50\% and 80\%. Generally, electrosprays transitioned from a high-efficiency droplet regime at $\dot{m} > 10^{-11}$\,kg/s, to an ion-droplet mixed regime with lower efficiency at $4\times10^{-12}$\,kg/s $<\dot{m} < 10^{-11}$\,kg/s, and finally to a purely ionic regime for $\dot{m} < 4\times10^{-12}$\,kg/s. The latter was only achieved for BMI-TCM. The $I_\text{sp}$ increased and thrust decreased as the mass flow rate was decreased.

Our study also identified significant discrepancies between the mass flow rate measured directly via the flow meter and indirectly through TOF at low flow rates. Consequently, TOF measurements may be unreliable for determining the mass flow rate, and therefore the specific impulse, of electrosprays operating in the ion-droplet mixed regime and in the purely ionic regime. 

\section*{Acknowledgments}

This work was funded by the Air Force Office of Scientific Research, award number FA9550-21-1-0200, the Defense Advanced Research Projects Agency, award number HR00112490392, and fellowships from the Balsells Foundation and the William and Ida Melucci Space Exploration \& Technology Fellowship. M. Caballero-P\'erez would like to sincerely thank Marco Magnani, Kaartikey Misra, and Juan Fern\'andez de la Mora for their insightful discussions.

    \section*{Supplemental material}
    \subsection*{Uncertainty in measurements}
    The total mass flow rate measurements in this study were calculated using $\dot{m} = \rho P/R_H$ and Eq. \eqref{eq:hyd_res} for finding the hydraulic resistance of the emitter capillaries $R_H$. The uncertainty of this measurement is:

    \begin{equation}
        \sigma_{\dot{m}} = \dot{m} \sqrt{ (\sigma_P/P)^2 + (\sigma_{P\text{h}}/P)^2 + (\sigma_{\Delta x}/\Delta x)^2 + (2 \sigma_\phi/\phi)^2 + (\sigma_\mu/\mu)^2 + (\sigma_\rho/\rho)^2}
    \end{equation}

    The first source is the uncertainty due to the liquid feeding pressure. The pressure at the propellant reservoir was controlled automatically and oscillated $\pm 1\,$mmHg around the set point. The lowest pressure used during experiments was 50\,mmHg, hence this source of error was always lower than 4\%. The second source of error is that associated with a small pressure head component $P_{\text{h}} = \rho g_0 h$ due to the liquid end being at a height difference $h$ with the emitter. The pressure head contributes to about 1\,mmHg per cm of height difference. Care was taken to ensure that this source of error was negligible. The third source of error is associated with the increment of the distance traveled by the propellant interface at the flow meter to determine the hydraulic resistance. The ruler that was used had increments of 0.8\,mm and marking lines 0.12\,mm thick, making the uncertainty in the distance measured $\sigma_{\Delta x} \sim 0.12$\,mm. The liquid interface was allowed to travel a sufficiently large $\Delta x$ ($>8$\,mm) for this source of uncertainty to become low. The fourth main source of uncertainty was that of the inner diameter of the flow meter $\phi$. For the 100 and \SI{200}{\micro\meter} diameter capillaries, the uncertainty provided by the manufacturer (Molex Polymicro) is 4 and \SI{6}{\micro\meter} in diameter, resulting in relative errors of 8\% and 6\% on the total mass flow rate, respectively. This is the largest contributor to uncertainty. Finally, there are uncertainties in viscosity and density due to temperature changes throughout the experiments. Although the lab is temperature-controlled and varied at most 1$^\circ$C throughout the experiments, the ionic liquids tested change around 4 to 5\% in viscosity per degree Celsius, so this uncertainty component is also relatively large. The change in density, however, is small. Overall, this resulted in a relative uncertainty in the total mass flow rate of about 10\% for all measurements. 

    The uncertainty in TOF thrust $T'$ is computed as:

    \begin{equation}
        \sigma_{T'} = T' \sqrt{ \left(\frac{\sigma_I}{\mu_I \sqrt{n_I}}\right)^2 + \left(\frac{\sigma_\Lambda}{\mu_\Lambda \sqrt{n_T}}\right)^2  }
        \label{eq:unc_thrust}
    \end{equation}

    The uncertainty of the thrust at 10\,kV ($T'_a$) is $\sigma_{T'} \sqrt{V_a/V}$, where $V$ is the emitter voltage and $V_a=10$\,kV. The first term in Eq. \eqref{eq:unc_thrust} is the uncertainty of the emitted current. The standard error in its measurement ($\sigma_I/\sqrt{n_I}$) is used, where $\sigma_I$ is the standard deviation of the current measurements, $\mu_I$ is the average, and $n_I$ is the number of sample points. The second term is the uncertainty associated with the time-of-flight signal integral $\Lambda = \int_0^\infty (1-{F}_\tau(t))dt$. The standard error is also used as its uncertainty. 

    The uncertainty in the specific impulse at 10\,kV ($I'_{\text{sp},a}$) is: 

    \begin{equation}
        \sigma_{\text{Isp}} = I'_{\text{sp},a} \sqrt{(\sigma_{T'}/T')^2 + (\sigma_{\dot{m}}/\dot{m})^2}
    \end{equation}

    And for the efficiency, it is:

    \begin{equation}
        \sigma_{\eta'} = \eta' \sqrt{\left(\frac{2 \sigma_\Lambda}{\mu_\Lambda \sqrt{n_T}}\right)^2  +\left(\frac{\sigma_\Phi}{\mu_\Phi \sqrt{n_T}}\right)^2 }
    \end{equation}

    Here, $\Phi = \int_0^\infty t(1-{F}_\tau(t))dt$, $\mu_\Phi$ is the mean of $\Phi$ from $n_\Phi$ samples of time-of-flight traces. The uncertainty in the ratio $r=\dot{m}'/\dot{m}$ between TOF mass flow rate and total mass flow rate is:

    \begin{equation}
        \sigma_r = r \sqrt{\left(\frac{\sigma_\Phi}{\mu_\Phi \sqrt{n_\Phi}}\right)^2  + \left(\frac{\sigma_I}{\mu_I \sqrt{n_I}}\right)^2 + \left(\frac{ \sigma_{\dot{m}}}{\dot{m}}\right)^2 }
    \end{equation}

\bibliography{main}

\end{document}